\documentclass[11pt,a4paper]{article}
\usepackage{jcappub}

\usepackage{url}
\usepackage{lscape}

\def\beq{\begin{equation}}
\def\eeq{\end{equation}}
\def\alwaysmath#1{{\ifmmode{#1}\else{$#1$}\fi}}

\title{
Mapping systematic errors in helium abundance determinations using Markov Chain Monte Carlo
}

\author[a]{Erik Aver}
\author[b]{Keith~A.~Olive}
\author[c]{Evan~D.~Skillman}

\affiliation[a]{School of Physics and Astronomy, University of Minnesota, \\
116 Church St. SE, Minneapolis, MN 55455}
\emailAdd{aver@physics.umn.edu}
\affiliation[b]{William I. Fine Theoretical Physics Institute, University of Minnesota, \\
116 Church St. SE, Minneapolis, MN 55455}
\emailAdd{olive@umn.edu}
\affiliation[c]{Astronomy Department, University of Minnesota, \\
116 Church St. SE, Minneapolis, MN 55455}
\emailAdd{skillman@astro.umn.edu}

\abstract{
Monte Carlo techniques have been used to evaluate the statistical and systematic uncertainties in the helium abundances derived from extragalactic H~II regions.  The helium abundance is sensitive to several physical parameters associated with the H~II region. In this work, we introduce Markov Chain Monte Carlo (MCMC) methods to efficiently explore the parameter space and determine the helium abundance, the physical parameters, and the uncertainties derived from observations of metal poor nebulae. Experiments with synthetic data show that the MCMC method is superior to previous implementations (based on flux perturbation) in that it is not affected by biases due to non-physical parameter space.  The MCMC analysis allows a detailed exploration of degeneracies, and, in particular, a false minimum that occurs at large values of optical depth in the He~I emission lines.  We demonstrate that introducing the electron temperature derived from the [O~III] emission lines as a prior, in a very conservative manner, produces negligible bias and effectively eliminates the false minima occurring at large optical depth.  We perform a frequentist analysis on data from several ``high quality'' systems.  Likelihood plots illustrate degeneracies, asymmetries, and limits of the determination.  In agreement with previous work, we find relatively large systematic errors, limiting the precision of the primordial helium abundance for currently available spectra.
}

\keywords{}

\arxivnumber{}

\begin{document}

\begin{flushright}UMN-TH-2929/10\\FTPI-MINN-10/35\\
December 2010\end{flushright}
\vskip -0.68in

\maketitle
\flushbottom

\section{Introduction}

Standard big bang nucleosynthesis (SBBN) using the baryon density determined by WMAP \citep{wmap,wmap10} predicts the initial abundances of D, $^{3}$He, $^{4}$He, and $^{7}$Li \citep{cfo,coc,coc2,cyburt,coc3,cuoco,serp,cfo5}, allowing one to probe the early universe at redshifts of order $10^{10}$ \citep{wssok,osw,fs}.  Therefore, the observed abundances provide a valuable check on the theory of SBBN, its concordance with the measurements of the microwave background radiation, and the content and interactions of the universe during the period of BBN \citep{MM93,sar,cfos}.  To test these predictions, the observed abundances must be determined with relatively high precision.  Because of the logarithmic relationship between the baryon to photon ratio, $\eta$, and the primordial helium abundance, Y$_{p}$, the uncertainty of Y$_{p}$ must be $<1\%$ to meaningfully test the theory. The 7-year WMAP value for $\eta$ is $(6.19 \pm 0.15) \times 10^{-10}$, \citet{wmap10}.  For comparison, the SBBN calculation of \citet{cfo5}, assuming the WMAP $\eta$ and a neutron mean life of $885.7 \pm 0.8$ s \citep{rpp}, yields $Y_p = 0.2487 \pm 0.0002$, a relative uncertainty of only 0.08\%.

The determination of Y$_{p}$ is facilitated through low metallicity H~II regions in dwarf galaxies.  By fitting the helium abundance versus metallicity, one can extrapolate back to very low metallicity, corresponding to the primordial helium abundance \citep{ptp74}.  The oxygen to hydrogen ratio, O/H, commonly serves as a proxy for metallicity.  Though this area of research has benefited from three decades of development, the determinations of Y$_{p}$ have suffered from significant differences between the results.  The difficulties in calculating an accurate and precise measure of the primordial helium abundance are well established \citep{os01,os04,its07}. Here, we introduce a new method based on Markov Chain Monte Carlo (MCMC) techniques.

Observations of the helium to hydrogen emission line ratio from extragalactic H~II regions provide a measure of the helium to hydrogen ratio, y$^{+}={n(He~II) \over n(H~II)}$.  Correspondingly, the statistical measurement errors in the helium and hydrogen emission line fluxes contribute to the uncertainty on y$^{+}$.  Unfortunately, this calculation of y$^{+}$ and its uncertainty are complicated by a myriad of systematic effects.  Interstellar reddening, underlying stellar absorption, radiative transfer, and collisional corrections alter the observed flux, complicating the measurement of y$^{+}$, and amplifying the uncertainty.  The photons are scattered by dust on their journey (interstellar reddening).  The stellar continuum juxtaposes absorption features under nebular emission lines (helium and hydrogen underlying absorption).  The H~II region itself absorbs and re-emits photons (radiative transfer); both recombination and collisional excitation contribute to the emission (collisional corrections for helium and hydrogen).  None of these processes can be directly measured and, therefore, cannot be removed independent of the observed emission lines and theoretical models.  As a result, the uncertainty on  y$^{+}$ must reflect the presence of, and lack of certainty regarding, these systematic effects.  Determining y$^{+}$ in conjunction with the robust estimation of the model parameters used in correcting for the listed systematic effects requires ``high quality'' spectra.  This desired confidence weighs against the need for larger sample sizes to decrease the uncertainty on Y$_p$ (and $dY/dZ$).

The importance of Monte Carlo techniques was demonstrated in a ``self-consistent'' analysis method, stemming from the work of \citet{itl94} and \citet{ppr00}, for determining the nebular helium abundance based upon six helium and three hydrogen lines \cite{os01,os04}.   In preceding work \cite{AOS}, hereafter AOS, we updated and extended the physical model and integrated the helium and hydrogen calculations with the goals of improving accuracy and removing assumptions.  The focus of the current paper is the exploration of a new technique based on a Markov Chain Monte Carlo (MCMC) analysis and departs from the ``self-consistent'' method.  Rather than fitting the parametric inputs to a helium abundance, the frequentist approach developed here builds a global likelihood function for all parameters including the helium abundance.

As we will demonstrate, the MCMC method is statistically superior to previous efforts, it is more direct and transparent, and it maintains efficiency.  Of primary benefit, the results are more rigorous:  the solution remains unbiased by the procedure, the uncertainty captures the confidence of the model and measurements, visualization of the parameter space topology reveals the reliability of the determination, and spectra failing to resolve their physical environments are identified.

Section \ref{ParameterSimulation} discusses the determination of parameter uncertainties and details the differences in our approach between AOS and this work.  The utility of Markov Chains and the computational implementation are described in \S \ref{MCMC}.  In \S \ref{Synthetic}, we describe tests using synthetic data to illustrate the method and its utility, with particular emphasis on secondary minima and the incorporation of a temperature prior.  MCMC is implemented in analyzing the dataset used in AOS in \S \ref{Galaxies}, and, in \S \ref{Results}, Y$_p$ is determined.  Finally, \S \ref{Conclusion} offers a discussion of the exploration and results as well as the next steps in better determining the primordial helium abundance.

\section{Parameter simulation} \label{ParameterSimulation}

The purpose of the Monte Carlo is to generate a set of parameters with their $\chi^{2}$ values.  Improving upon the previous efforts, a new technique, MCMC (see \S \ref{MCMC}), enables simultaneous Monte Carlo over all model parameters including the He abundance.  A full frequentist sampling of the parameter space involving eight input parameters would be computationally prohibitive. However, using MCMC allows one to judicially sample regions of parameter space with relatively low $\chi^2$.  In this way, we can construct the likelihood function and determine the best fit point in the multidimensional parameter space along with their associated
uncertainties.  Indeed, the uncertainties are the primary focus of this work. 

The Monte Carlo approach of ref.~\cite{os04} and AOS took each set of measured fluxes and built a Gaussian distributed dataset of fluxes based upon their measurement uncertainty.  For each of a 1000 such datasets, a best-fit solution was found for the helium abundance as well as the physical input parameters using the ``self-consistent" method which determines the set of input parameters with a $\chi^2$ based on the derived helium abundance from each of six helium emission lines.  The final result was computed from the average and standard deviation of the set of solutions.  
Using the fluctuation of the minimum is, however, not a direct measure of the $\chi^{2}$'s parameter dependence.  Furthermore, it is also not as robust as desired.  Each solution was restricted to physically meaningful parameter space (e.g., positive densities), potentially biasing the solution.  Additionally, as was manifested in AOS and will be discussed further in \S \ref{Synthetic}, $\chi^{2}$ functions lacking a well constrained temperature and density can produce unlikely high density and low temperature solutions that greatly skew the results.  Ultimately, these considerations, tempered by the required computational efficiency, motivate this work.

The $\chi^{2}$ function defined here, and used for parameter fitting, is modified from that used in previous work.  Rather than defining y$^{+}$ implicitly, as the average of six individual line abundances, and minimizing the deviation between the lines, y$^{+}$ is demoted to an input parameter, no different than the others (e.g., temperature and density).  Instead, here, we use all of the input parameters (described below) and calculate synthetic fluxes which are then compared to observed flux, weighted by the observed uncertainty, allowing for a more standard definition of $\chi^{2}$, 
\beq
\chi^2 = \sum_{\lambda} {(\frac{F(\lambda)}{F(H\beta)} - {\frac{F(\lambda)}{F(H\beta)}}_{meas})^2 \over \sigma(\lambda)^2},
\label{eq:X2}
\eeq
where the He flux at each wavelength $\lambda$ relative to the flux in $H\beta$ is given by
\beq
\frac{F(\lambda)}{F(H\beta)}= y^{+}\frac{E(\lambda)}{E(H\beta)}{\frac{W(H\beta)+a_{H}(H\beta)}{W(H\beta)} \over \frac{W(\lambda)+a_{He}(\lambda)}{W(\lambda)}}{f_{\tau}(\lambda)}\frac{1+\frac{C}{R}(\lambda)}{1+\frac{C}{R}(H\beta)}10^{-f(\lambda)C(H\beta)}.
\label{eq:F_He_EW}
\eeq
The  $\chi^{2}$ in eq. \ref{eq:X2} runs over He and H lines, and the ratio of H fluxes is defined analogously, 
\beq
\frac{F(\lambda)}{F(H\beta)}= \frac{E(\lambda)}{E(H\beta)}{\frac{W(H\beta)+a_{H}(H\beta)}{W(H\beta)} \over \frac{W(\lambda)+a_{H}(\lambda)}{W(\lambda)}}\frac{1+\frac{C}{R}(\lambda)}{1+\frac{C}{R}(H\beta)}10^{-f(\lambda)C(H\beta)}.
\label{eq:F_H_EW}
\eeq
For the above flux equations, six measured helium emission line fluxes ($\lambda$3889, 4026, 4471, 5876, 6678, and 7065) and three hydrogen emission line fluxes ($H\alpha$, $H\gamma$, $H\delta$), each relative to $H\beta$ ($\frac{F(\lambda)}{F(H\beta)}$), along with their equivalent widths ($W(\lambda)$) are used.  The predicted model fluxes are calculated from an input value of y$^{+}$ and emissivity ratio of H$\beta$ to the helium or hydrogen line, $\frac{E(H\beta)}{E(\lambda)}$, with corrections made for reddening (C(H$\beta$)), underlying absorption (a$_{H}$ \& a$_{He}$), collisional enhancement, and radiative transfer.  The optical depth function, \textit{$f_{\tau}$}, and collisional to recombination emission ratio, \textit{$\frac{C}{R}$}, are both temperature (T) and density (n$_{e}$) dependent (the emissivities are also temperature dependent).  Additionally, the hydrogen collisional emission depends on the neutral to ionized hydrogen ratio ($\xi$).  Therefore, there are a total of eight model parameters (y$^{+}$, n$_{e}$, a$_{He}$, $\tau$, T, C(H$\beta$), a$_{H}$, $\xi$).  The physical model itself, the equations relating the abundance and correction parameters to the flux, is unchanged from AOS, and incorporates all of the effects explored therein (updated emissivities, wavelength dependent underlying absorption, and neutral hydrogen collisional emission).  

We note here that the equivalent width of an emission line is not independent of the flux.  Each flux is related to its equivalent width through the flux of the continuum at that wavelength, h($\lambda$), as
\beq
F(\lambda) = W(\lambda)h(\lambda).
\eeq
As a result, if the model parameters generate a lower flux, the equivalent width should be lowered correspondingly.  The continuum flux is constrained to remain constant such that the ratio $\frac{h(\lambda)}{h(H\beta)}$ is determined from the measured flux ratio and equivalent widths,
\beq
\frac{h(\lambda)}{h(H\beta)} = {\frac{F(\lambda)}{F(H\beta)}}_{meas} \frac{W(H\beta)_{meas}}{W(\lambda)_{meas}}.
\eeq
Eq. \ref{eq:F_He_EW} can therefore be rewritten to remove $W(\lambda)$ entirely and solve for a consistent emission line ratio to H($\beta$).  This yields a simplified He flux ratio,
\beq
\frac{F(\lambda)}{F(H\beta)}= y^{+}\frac{E(\lambda)}{E(H\beta)}{f_{\tau}(\lambda)}\frac{1+\frac{C}{R}(\lambda)}{1+\frac{C}{R}(H\beta)}10^{-f(\lambda)C(H\beta)}\frac{W(H\beta)+a_{H}(H\beta)}{W(H\beta)}-\frac{a_{He}(\lambda)}{W(H\beta)}\frac{h(\lambda)}{h(H\beta)},
\label{eq:F_He}
\eeq
and H flux ratio,
\beq
\frac{F(\lambda)}{F(H\beta)}= \frac{E(\lambda)}{E(H\beta)}\frac{1+\frac{C}{R}(\lambda)}{1+\frac{C}{R}(H\beta)}10^{-f(\lambda)C(H\beta)}\frac{W(H\beta)+a_{H}(H\beta)}{W(H\beta)}-\frac{a_{H}(\lambda)}{W(H\beta)}\frac{h(\lambda)}{h(H\beta)}.
\label{eq:F_H}
\eeq 
Because it is this ratio of fluxes that is calculated, $W(H\beta)$ cannot be removed from the equation.  The treatment of this measured quantity (equivalent to the H$\beta$ flux measurement) will be discussed in the next section (\S \ref{MCMC}).


To reduce bias in the solutions of AOS, the emissivities of \citep[][PFM]{pfm07} were refit with a stronger density dependence to protect the $\chi^2$ minimizations from running away to 
unphysical regions of the parameter space.  As mentioned above, the method of AOS was susceptible to the skewing of the final average.  Because at large densities the density dependence of the PFM emissivities is negligible, extreme outliers with densities increasing nearly without bound occurred, rendering the final result meaningless.  Refitting the emissivities to the form of \citet{bss99} rectified this behavior, while preserving the accuracy of the PFM equations in the region of interest.  However, the approach of this work, direct Monte Carlo over the parameters, is not susceptible to the previously exhibited biasing.  A poor constraint on the density will be manifested in the $\chi^{2}$ function.  This will affect the uncertainty in a transparent way, reflecting the quality of the determination.  The solution, the minimum $\chi^{2}$, will be unaffected, and any degeneracy in the solution, i.e., multiple local minima, will be apparent.  As a result, the unmodified PFM emissivities \citep{pfm07} are used in this work.

\section{Markov Chain Monte Carlo} \label{MCMC}

The Markov Chain Monte Carlo (MCMC) method is an algorithmic procedure for sampling from a statistical distribution \citep{mar,met}.  Therefore, the sequence of points in the parameter space reconstructs the target distribution, here the likelihood distribution ($\mathcal{L}$).  The real value of MCMC is the judicious choice of sampling such that the density of samples is greatest around the best-fit point and increasingly sparse in increasingly unlikely regions in parameter space.  As a result, computational time is not wasted in exploring poor model solutions, while full exploration of complex topology is still facilitated.  In brief, the algorithm is efficient, with computational time scaling roughly linearly with the number of parameters.  This is in sharp contrast to the exponential growth in time cost encountered with gridding.  For example, if only 100 points in each parameter were taken, then for eight parameters, a total of ${(10^2)}^8$ evaluations would be needed.  Typically, the results of this work are based on $10^6-10^7$ points.

CosmoMC\footnote{\url{http://cosmologist.info/cosmomc/}} is a freely available Fortran package for MCMC, designed originally for WMAP parameter extraction, but easily modified for generic sampling.  It allows for a variety of sampling algorithms, but this work makes use of the classic, and widely applicable, Metropolis-Hastings algorithm \citep{met,has}. We must also choose a proposal function which is a distribution function determining the sizes of the steps to take in each parameter.  The most common choice, and the one used here, is an $n$-dimensional Gaussian distribution, where $n$ is the number of input parameters.  For a symmetric proposal function, with $N$ points in the Markov chain:

\begin{minipage}{\textwidth}
\begin{itemize}
 \item Select an initial parameter vector, $\vec{x}_{0}$
 \item For $i=0$ to $N-1$:
\begin{enumerate}
 \item Randomly choose $\Delta \vec{x}$ from a proposal function
 \item $\vec{x}_{i+1} = \vec{x}_{i} + \Delta \vec{x}$
 \item $r=\frac{\mathcal{L}(x_{i+1})}{\mathcal{L}(x_{i})}$
 \item Accept $\vec{x}_{i+1}$ with probability $r$ (if $r>1$, $r=1$)
 \item Otherwise $\vec{x}_{i+1} = \vec{x}_{i}$
\end{enumerate}
\end{itemize}
\end{minipage}

In effect, a Gaussian distributed random walk is used; whereby, points that are of higher likelihood are always accepted, and points of lower likelihood are accepted proportional to the ratio of the proposed to current likelihood.  For this work, the likelihood is simply, 
\beq
\mathcal{L}=\exp(-\chi^2/2),
\eeq
with $\chi^{2}$ given by eq. \ref{eq:X2}.  After a `burn-in' period, the points will generate a Markov Chain converging to the target distribution.  The most important tuning comes in picking the proposal function step size, the Gaussian width.  If this is too large, then the acceptance fraction will be very small, and the points will be clumped with large separations.  If the width is too small, the exploration of the full distribution will be very slow, and failure to explore all minima becomes of increasing concern.  Roughly, a proposal width similar to the target distribution width, for each parameter, is the most efficient.  Furthermore, proposal widths leading to an acceptance fraction of $\sim$25\% have been shown to be optimally efficient \citep{hc,gel}.

One complication in this analysis comes from the emission line equivalent width of H$\beta$, $W(H\beta)$.  It is used in making the corrections for underlying absorption, and is, therefore, needed to calculate the predicted flux, even though it is itself a measured quantity.  To treat this, $W(H\beta)$ is encoded as a nuisance parameter.  For each MCMC point, a new Gaussian distributed equivalent width is chosen based upon the measured value and error.  A corresponding term is then also added into the $\chi^{2}$,
\beq
\chi^2_{W} = {(W(H\beta) - W(H\beta)_{meas})^2 \over \sigma(H\beta)^2}.
\eeq

Finally, the use of priors is straightforward in this MCMC analysis.  The restriction of the parameters to physically meaningful values, such as densities greater than zero, just means that proposal points in unphysical regions are automatically rejected.  Additional measurement information is incorporated directly into the likelihood in the form of an additional $\chi^{2}$ contribution.  For example, a temperature characterizing the H~II region and derived from other emission lines could be incorporated as,
\beq
\mathcal{L}=\exp(-\chi^2/2)\exp(-(T-T_{meas})^2/2\sigma^2).
\eeq
Consequently, points with temperatures diverging significantly from T$_{meas}$ would be disfavored by their low likelihood, and the uncertainty on the best-fit temperature, and potentially all other parameters, would be reduced.

\section{Synthetic exploration} \label{Synthetic}

\subsection{Improved analysis with MCMC} \label{Syn1}

\begin{figure}[ht!]
\centering  
\resizebox{\textwidth}{!}{\includegraphics{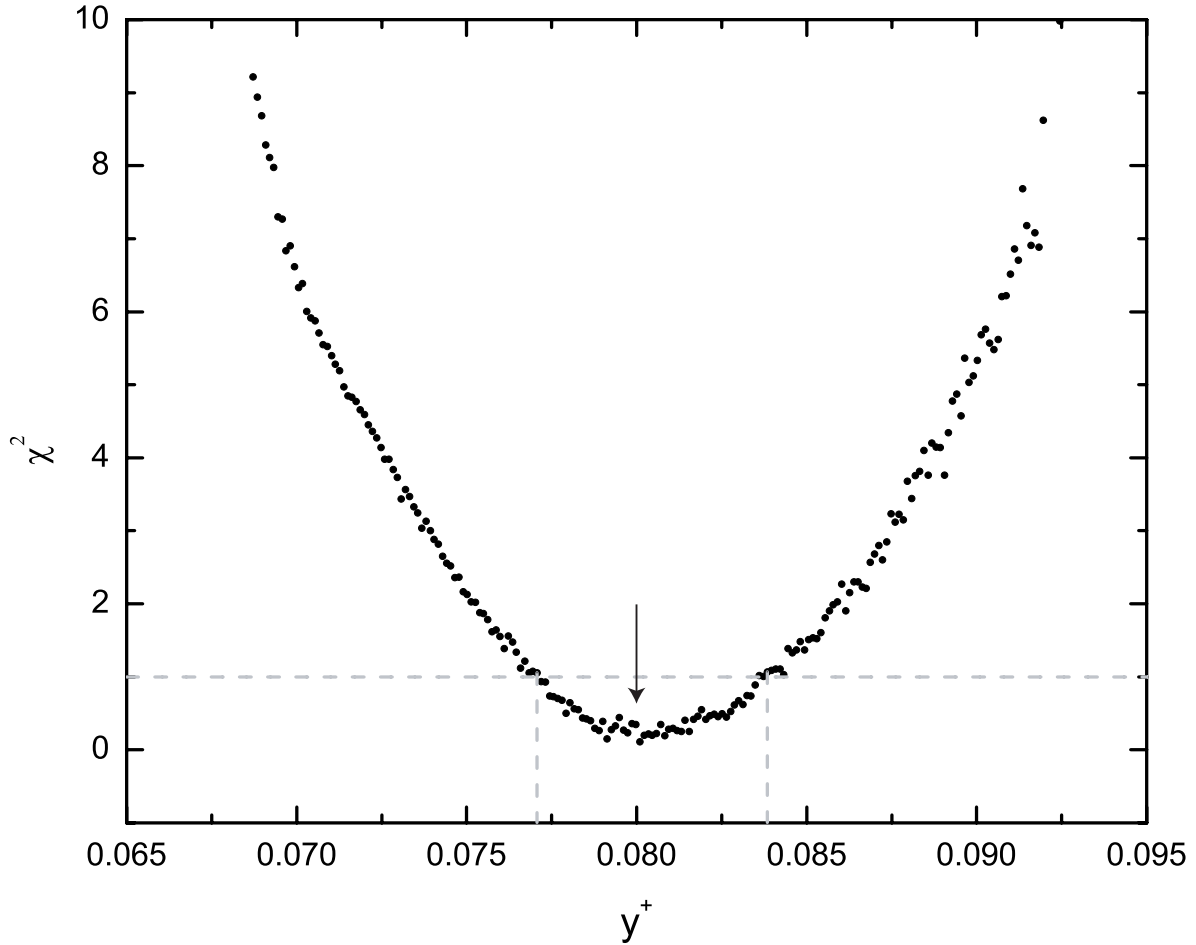}}
\caption{
$\chi^{2}$ versus abundance for synthetic data with model parameters, $y^{+}=0.08$, $n_{e} = 100~cm^{-3}$, $a_{He} = 1.0$ \AA, $\tau = 0.2$, $T = 18,000$ K, $C(H\beta) = 0.1$, $a_{H} = 1.0$ \AA, and $\xi = 1.0 \times 10^{-4}$.  The 68\% confidence level is marked by the dashed lines ($\chi^{2}_{min} = 0.0$ for synthetic data).   The arrow denotes the input value.
}
\label{Syn-y}
\end{figure}

\begin{figure}[ht!]
\centering  
\resizebox{\textwidth}{!}{\includegraphics{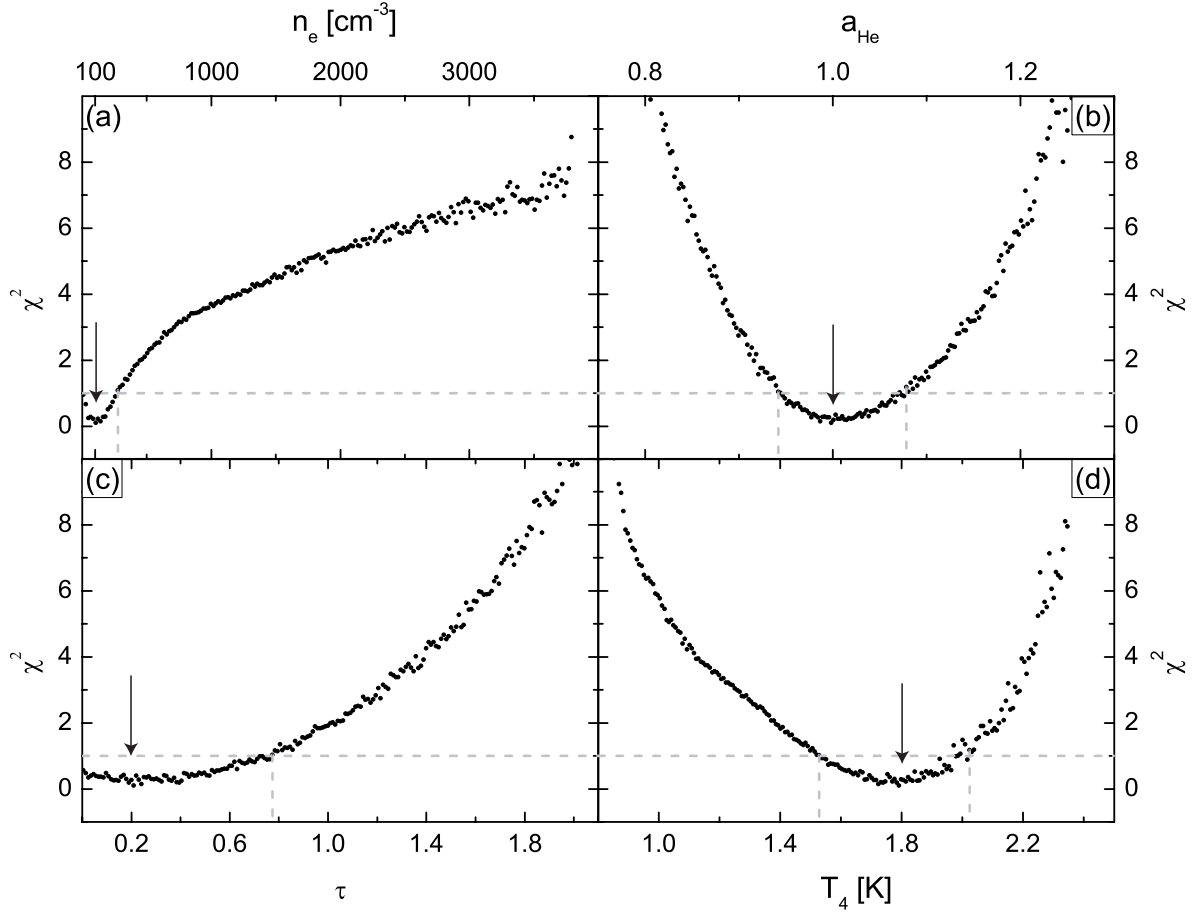}}
\caption{
Similar to figure \ref{Syn-y} with plots of $\chi^{2}$ versus density, helium absorption, optical depth, and temperature.
}
\label{Syn-He_4panel}
\end{figure}

\begin{figure}[ht!]
\centering  
\resizebox{\textwidth}{!}{\includegraphics{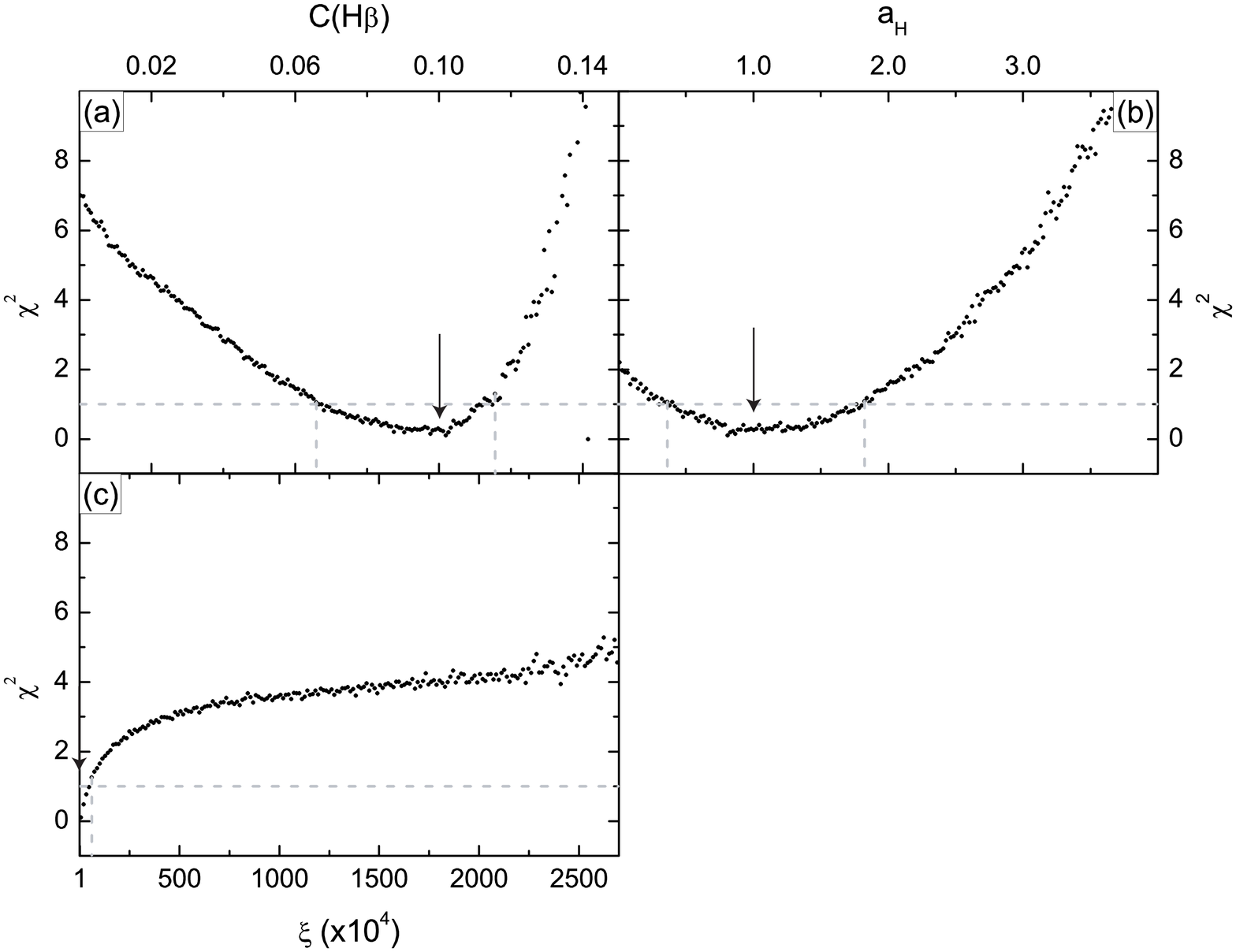}}
\caption{
Similar to figure \ref{Syn-y} with plots of $\chi^{2}$ versus reddening, hydrogen absorption, and neutral hydrogen fraction.
}
\label{Syn-H_3panel}
\end{figure}

Synthetic data are useful to demonstrate the utility and simplicity of the MCMC analysis.  Synthetic helium and hydrogen fluxes were generated using characteristic parameter values of  y$^{+}$ = 0.08, n$_{e}$ = 100 cm$^{-3}$, a$_{He}$ = 1.0 \AA, $\tau$ = 0.2, T = 18,000 K, C(H$\beta$) = 0.1, a$_{H}$ = 1.0 \AA, and $\xi = 1.0 \times 10^{-4}$.  Flux errors were 1\% for the hydrogen and 2\% for the helium lines with EW(H$\beta$) = 250 \AA.  Given the synthetic fluxes generated from these input parameters, we can apply the MCMC routine to determine the likelihood function across our 8-dimensional parameter space.  Figure \ref{Syn-y} shows the 1D marginalized likelihood for the helium abundance.  Each point shown represents a minimization of the $\chi^2$ function over variations of the other seven input parameters at a given (binned) value of y$^{+}$.  As one can see, the $\chi^2$ distribution nicely reproduces the input value of y$^{+}$ with the 68\% CL range indicated by the dashed lines.  Figures \ref{Syn-He_4panel} and \ref{Syn-H_3panel} show the other seven parameters.   The 68\%  CL ranges for all of the parameters are collected in table \ref{table:Synthetic}.

\begin{figure}[ht!]
\centering  
\resizebox{\textwidth}{!}{\includegraphics{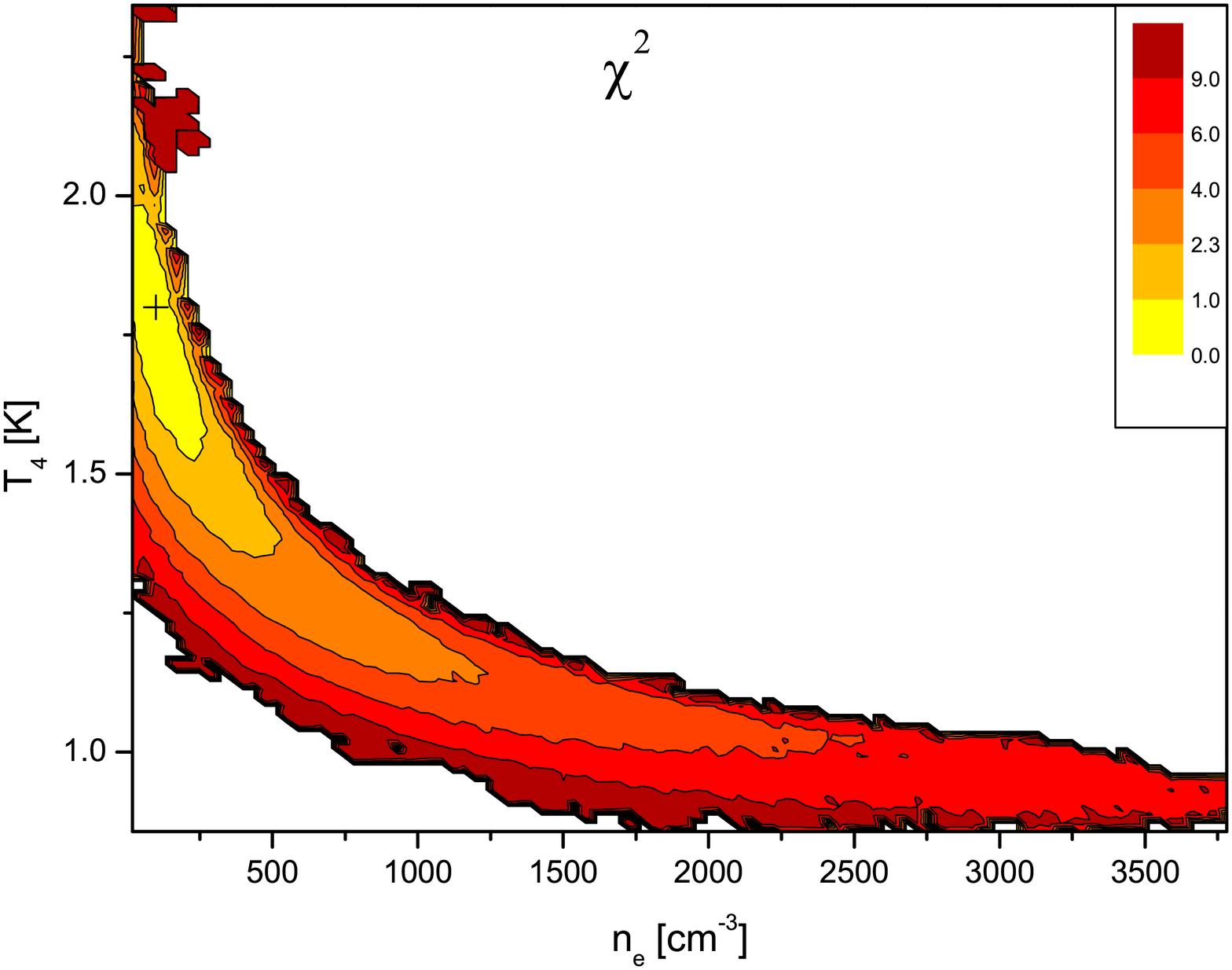}}
\caption{
Contour plot of $\chi^{2}$ versus density and temperature for the same synthetic model used in figures \ref{Syn-y}-\ref{Syn-H_3panel}.  Contours are $\Delta \chi^{2}$ = \{1, 2.3, 4, 6, 9\}.  The degeneracy between the parameters limits the precision of their determination and, therefore, of the helium abundance.
}
\label{Syn-D_T}
\end{figure}

The likelihood plots are revealing in several important ways.  First, the similarities between the abundance and temperature are not a coincidence (figures \ref{Syn-y} and \ref{Syn-He_4panel}d).  The temperature is the most important parameter in determining y$^{+}$ through the emissivity equations (see, e.g., AOS).  Also, because of a degeneracy in the determination, primarily between the 
temperature and the density (cf.\ AOS), the temperature is not well constrained and admits values 7,000 K below 
the input value at the 95\% level.  As shown in figure \ref{Syn-D_T}, there is a strong negative correlation between these parameters which allows a large variation in their values without rapidly raising $\chi^{2}$.  The 95\% levels, defined by the $\chi^{2}=4$ contour, allow densities reaching 1250 cm$^{-3}$ and temperatures ranging between 11,000 K and 22,000 K, and result in the abundances spanning 0.072 to 0.089.  

Second, the neutral to ionized hydrogen ratio (figure \ref{Syn-H_3panel}c) is unconstrained at large values, exhibiting asymptotic behavior at $\chi^{2}=4$.  The collisional relative to recombination contribution to the hydrogen emission depends exponentially on the temperature (see AOS), and as a result, the low temperatures arising out of the temperature-density trade-off permit nearly any value of the neutral hydrogen fraction, including unphysical ones.  In the previous approach, this behavior lead to some of the flux perturbed solutions with very large neutral hydrogen fractions.  This greatly biases the final averaged result, $\xi = 27.24 \times 10^{-4}$, away from the input value of 1.0 $\times 10^{-4}$ (table \ref{table:Synthetic}).  This biasing is exacerbated by the restriction of solutions to physical values, prohibiting negative solutions within the dataset.  Such restrictions are physically meaningful and necessary but, because of the statistical approach, impact the final result detrimentally.  In less pronounced fashion than for the neutral hydrogen fraction, any parameter with an unperturbed best-fit point near zero suffers from this effect in the analysis of AOS.  The optical depth, $\tau$, with a generating value of 0.2, exhibits this biasing clearly, returning 0.42.  For the singular solution of the new MCMC analysis, the restriction to physical values is correct and carries no possibility of unintended bias.

The results based on the method of AOS are also shown in table \ref{table:Synthetic} and can be directly compared with the current results based on MCMC. 
The AOS result reproduces the input helium abundance fairly well (y$^{+}$ is 1.4\% high which is less than one third of the 4.3\% uncertainty, see table \ref{table:Synthetic}).  In contrast, using MCMC, the final result for the abundance is that of the true minimum, and any asymmetry in the likelihood impacts only the uncertainty.  The same is true for the other parameters as well.  

\begin{table}[ht!]
\centering
\vskip .1in
\begin{tabular}{lccc}
\hline\hline
				& Input	 & AOS     			& MCMC \\
\hline
He$^+$/H$^+$			& 0.08 	 & 0.08108 $\pm$ 0.00341 	& 0.07999 $^{+0.00385}_{-0.00292}$ \\
n$_e$				& 100.0  & 147.8 $\pm$    282.8		& 100.1 $^{+   175.4}_{-   100.1}$ \\
ABS(He~I)			& 1.0	 & 1.05 $\pm$  0.09		& 1.00 $^{+ 0.08}_{- 0.06}$ \\
$\tau$				& 0.2    & 0.42 $\pm$  0.41		& 0.20 $^{+ 0.57}_{- 0.20}$ \\
T$_e$				& 18,000 & 17,440 $\pm$ 2308 		& 17,999 $^{+2239}_{-2711}$ \\
C(H$\beta$)			& 0.1    & 0.08 $\pm$  0.03		& 0.10 $^{+ 0.02}_{- 0.03}$ \\
ABS(H~I)			& 1.0    & 1.35 $\pm$  0.84		& 1.00 $^{+ 0.82}_{- 0.64}$ \\
$\xi$ $\times$ 10$^4$   	& 1.0    & 27.24 $\pm$    187.27	& 0.98 $^{+    59.77}_{-     0.98}$ \\
\hline
\end{tabular}
\caption{Comparing Gaussian Distributed Fluxes (from AOS) and MCMC Analyses with Synthetic Data}
\label{table:Synthetic}
\end{table}

\subsection{Degeneracies at large values of optical depth} \label{Tau}

\begin{figure}
\centering  
\resizebox{\textwidth}{!}{\includegraphics{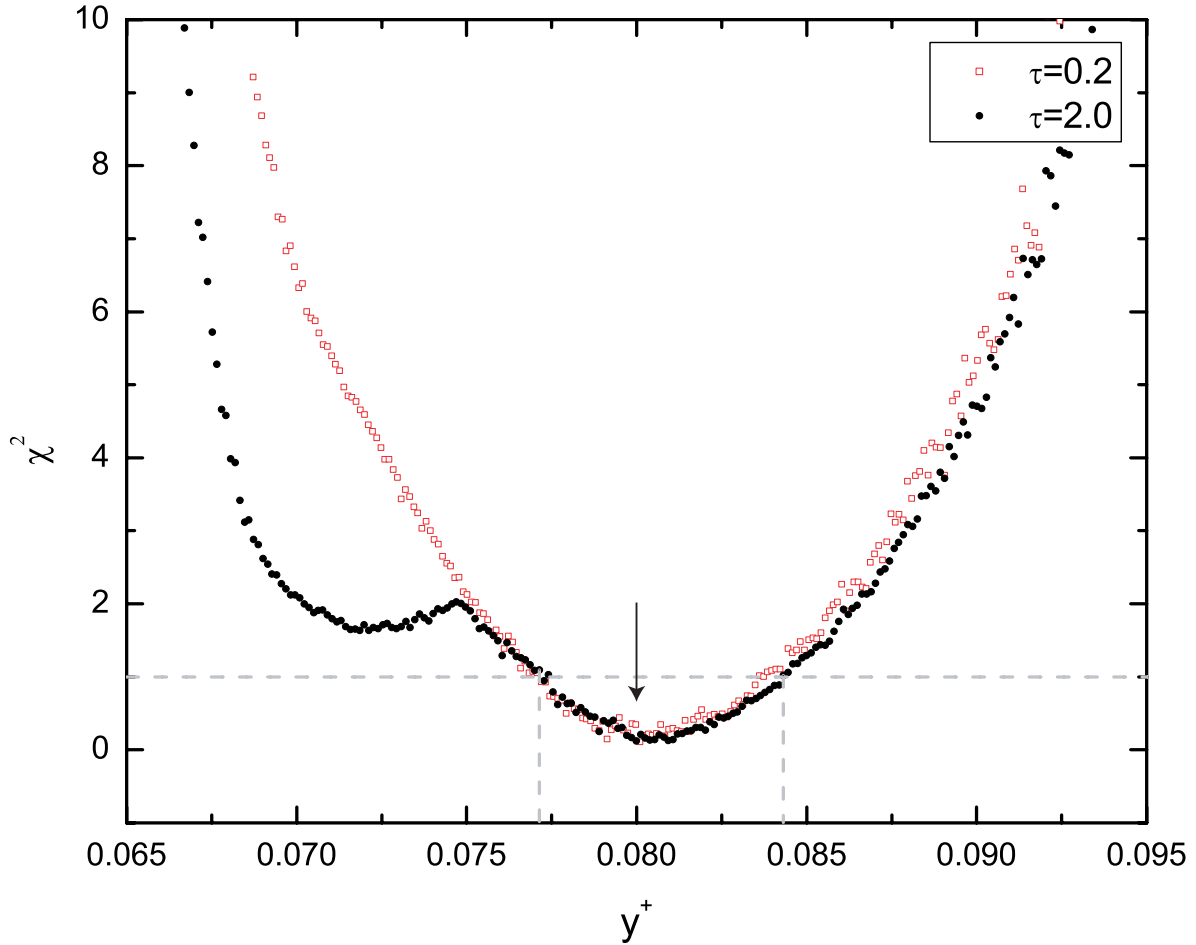}}
\caption{
$\chi^{2}$ versus abundance for synthetic data with model parameters, $y^{+}=0.08$, $n_{e} = 100~cm^{-3}$, $a_{He} = 1.0$ \AA, $T = 18,000$ K, $C(H\beta) = 0.1$, $a_{H} = 1.0$ \AA, and $\xi = 1.0 \times 10^{-4}$.  The only difference between the two sets is the value of the optical depth: $\tau = 0.2$ for the lighter, open squares and $\tau = 2.0$ for the darker, solid circles.  For the latter, the 68\% confidence level is marked by the dashed lines, and the input value is denoted by the arrow.  The larger optical depth allows a secondary minimum to develop at low abundance.  Such behavior highlights a deficiency in the model and mitigates the reliability of galaxies exhibiting large optical depth.
}
\label{Syn-Tau-y}
\end{figure}

An important feature emerges in synthetic testing at high optical depth.  Using the same generating values as in \S \ref{Synthetic}, except for $\tau$ = 2.0 instead of 0.2, results in a pronounced second minimum in the helium abundance (see figure \ref{Syn-Tau-y}).  For objects with significant radiative transfer and relatively large errors on He $\lambda$3889, the parameter space contains the flexibility to decrease $\tau$ abruptly from the best-fit value, at a correspondingly low temperature, low abundance, and high density, to decrease $\chi^2$.  Consequently, a second minimum is found in the MCMC analysis.  

The reason for this second minimum can be understood by examining figure 5 in ref.~\cite{os01}.
The He~I $\lambda$7065 emission line has a strong dependence on density (it is increased 
via collisional excitations from the meta-stable triplet 2s level).  However, at significant
values of optical depth, the He~I $\lambda$7065 emission line is also increased by absorptions
out of the meta-stable triplet 2s level which cascade down through this emission line.
Because the He~I $\lambda$3889, is also increased by via collisional excitations from the 
meta-stable triplet 2s level, but {\it decreased} by absorptions from that level, the two
lines can trade against each other, combined with the temperature-density degeneracy, to 
form the second, non-physical, minimum.

In figure \ref{Syn-Tau-y}, the second local minimum is near y$^{+}$ = 0.072, and correspondingly, T = 10,000 K, n$_{e}$ = 1500 cm$^{-3}$, and $\tau$ = 0.75.  Though it does not impact the 68\% CL, it lies well within the 95\% CL limit ($\chi^2$=1.6).  Results from the previous method (AOS) were in fact susceptible to being skewed by secondary minima.  Upon perturbation of the fluxes, this lower minimum could become the best-fit point for some of the datasets. Then, upon averaging, these outlier solutions bias the final result.  As already discussed, the MCMC solution is unaffected by asymmetries about the minimum; however, broad, deep secondary minimum will greatly, and artificially, increase the calculated uncertainty.  Furthermore, for real data, the presence of a second local minimum, with a similar $\chi^2$, raises doubts regarding which minimum represents the physical environment.  The object's reliability is clearly decreased, but, in the following section (\S \ref{TOIII}), the use of an additional observation of the temperature via the [O~III] emission lines to resolve this ambiguity is discussed.

This work uses the radiative transfer calculations of \citet{bss02} \citep[first used in][]{os04}.  These fits assume a spherical nebula with uniform density and temperature and no systematic velocity gradients.  However, H~II regions may be expanding -- the original work of \citet{rob68} models the regions as expanding with a constant velocity gradient.  Similarly, the effects of turbulence may be more complicated than as incorporated.  Any such deviations from the model will be more pronounced at larger optical depth.  Furthermore, He $\lambda$3889 is the primary line in determining $\tau$; yet, it is blended with H8, decreasing its reliability.  The combination of these concerns, in addition to the emergence of the secondary minimum, motivate a preference for objects with low optical depth.

\subsection{Using T(O~III) as a conservative  prior} \label{TOIII}

Adding an independent measurement of the temperature in the H~II region
has the potential to weaken the temperature-density degeneracy. 
The [O~III] $\lambda\lambda$4363, 4959, and 5007 emission lines provide such
and independent measurement, T(O~III), associated with the doubly ionized 
oxygen of the H~II region.  However, because
the temperature in the H~II region is not expected to be perfectly uniform, the
exponential sensitivity to temperature of the [O~III] emission lines biases the 
derived temperature to higher than average values.  
Thus, T(O~III) may not well characterize the average electron temperature 
over the entire He~II emission zone.   The magnitude, and thus importance, 
of this temperature difference is a matter of debate.  Recent work by \citet{ppl02} 
found that T(He~II) was always less than T(O~III), 
ranging between 6-10\% less (solved) and 3-11\% less (photo-ionization models).  
However, from comparisons of temperatures derived from the Balmer jump with 
temperatures derived from the [O~III] emission lines, Guseva et al. \citep{guseva06, guseva07} 
find no evidence for this temperature offset from studies of low
metallicity H~II regions.

To use T(O~III) directly as a measurement of the average temperature would 
break the temperature-density degeneracy (with resulting very small 
uncertainties), but risks biasing the derived values
of helium abundance to artificially high values in the presence of non-uniform
temperatures.  In fact, T(O~III) can be viewed as a limit to the upper bound on the 
average temperature in the H~II region, and as a consistency check to 
eliminate physically improbable low temperature solutions.
The goal is to use T(O~III) to constrain T(He~II) to physically meaningful values,
not to bias it.  Note that the expected differences between T(O~III) and T(He~II)
are much larger than the calculated error on T(O~III).   Therefore, in incorporating 
T(O~III) as a prior, this work takes $\sigma(T_{OIII})=0.2~T_{OIII}$, a very weak constraint.  
This will keep bias in the solution negligible, while employing the available 
additional temperature diagnostic to better isolate the physically relevant parameter space.  
In the case of likelihoods exhibiting a second minimum with a very low corresponding 
temperature (as demonstrated in \S \ref{Tau}), the T(O~III) prior rigorously and 
definitively rules out the unphysical region.  As a result, the parameter errors 
are then well defined, and the solution is unambiguous.  

\begin{figure}
\centering  
\resizebox{\textwidth}{!}{\includegraphics{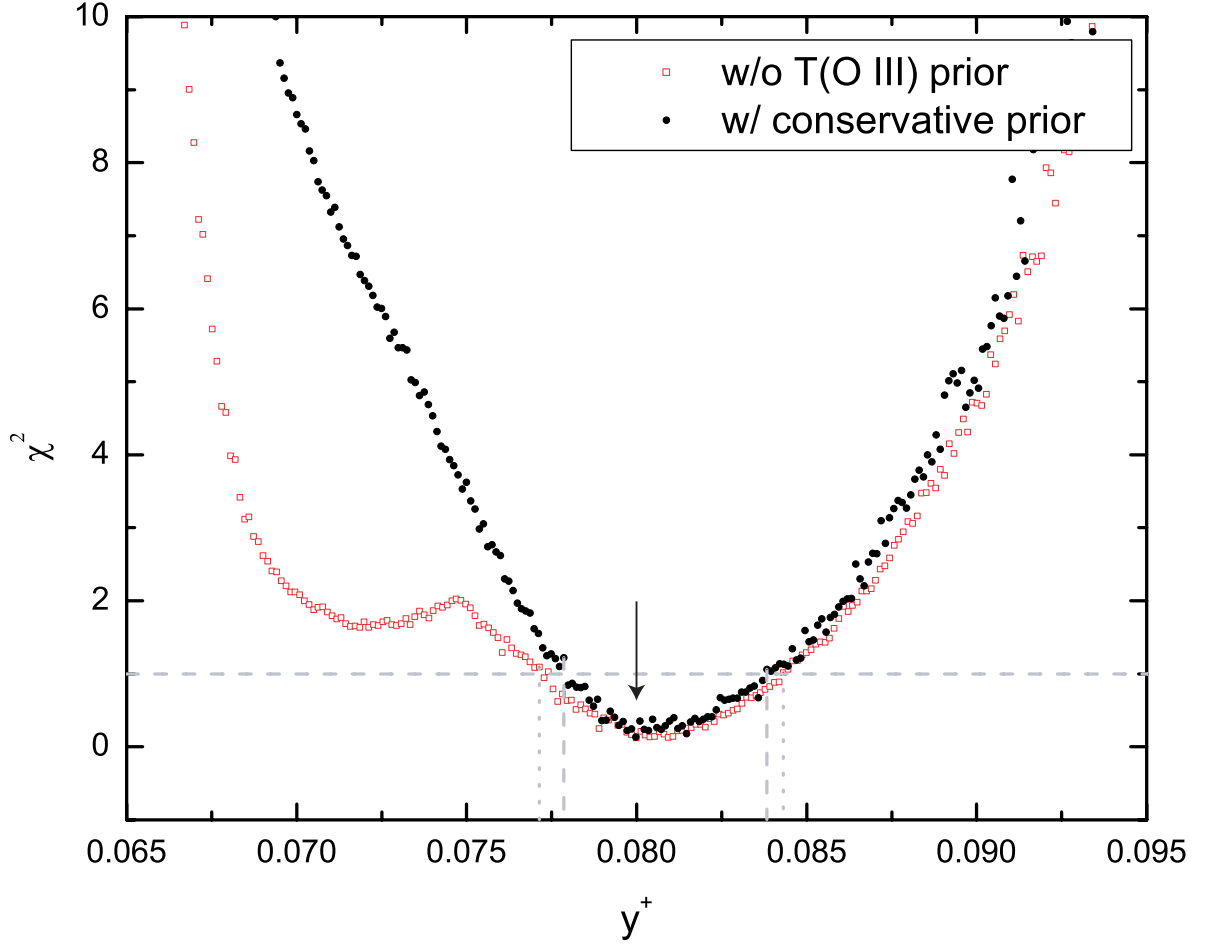}}
\caption{
$\chi^{2}$ versus abundance for synthetic data with model parameters, $y^{+}=0.08$, $n_{e} = 100~cm^{-3}$, $a_{He} = 1.0$ \AA, $\tau = 2.0$, $T = 18,000$ K, $C(H\beta) = 0.1$, $a_{H} = 1.0$ \AA, and $\xi = 1.0 \times 10^{-4}$.  The darker, solid circles are analyzed with $T(O~III) = 19,000 K$ included as a prior, while the lighter, open squares do not include the prior.  The 68\% confidence level is marked by the dashed lines, and the input value is denoted by the arrow.  Note that the minimum is not noticeably impacted by the prior.
}
\label{Syn-TOIII-y}
\end{figure}

\begin{figure}
\centering  
\resizebox{.95\textwidth}{!}{\includegraphics{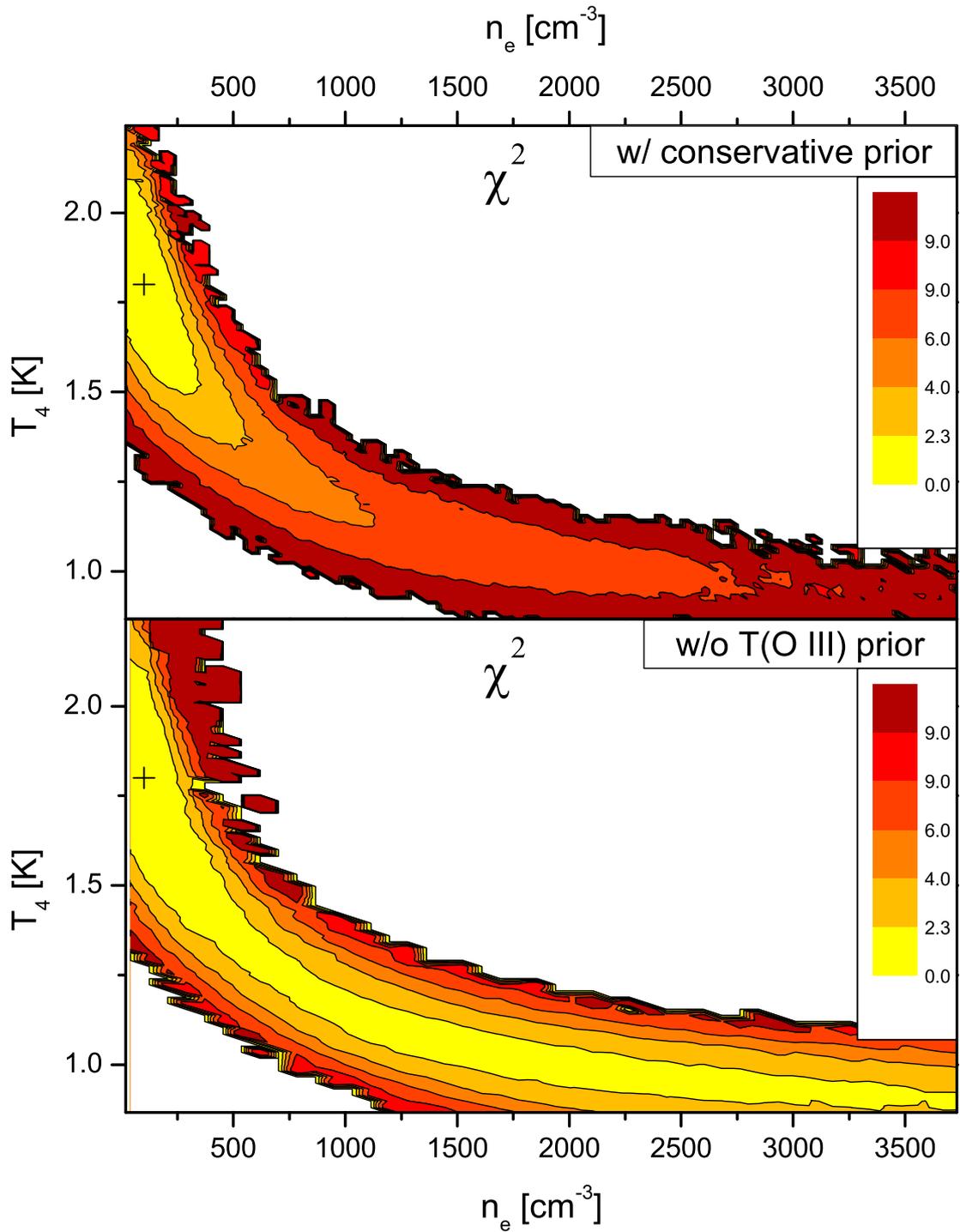}}
\caption{
Contour plot of $\chi^{2}$ versus density and temperature for the same synthetic model used in figure \ref{Syn-TOIII-y}.  Contours are $\Delta\chi^{2}$ = \{1, 2.3, 4, 6, 9\}.  The dramatic curtailing of the density and temperature variance is due solely to the inclusion of the weak T(O~III) prior -- with and without in the top and bottom panels, respectively.
}
\label{Syn-TauTOIII-D_T}
\end{figure}

Figure \ref{Syn-TOIII-y} illustrates the effect of using the conservative T(O~III) prior on the synthetic object's likelihood.  The low temperature local minimum is completely removed by using the T(O~III) measurement as a prior, even with the very gentle implementation chosen.  As expected, due to the temperature-density degeneracy, the increased constraint on temperature translates into a marked improvement in the derived density, as is evidenced in figure \ref{Syn-TauTOIII-D_T}.  The only concern is the possibility of bias in the best-fit solution.  For this example, T(O~III) was input as 19,000 K, 6\% higher than T(He~II).  Table \ref{table:Syn_TOIII} compares the minimum $\chi^2$ solution with and without the use of the prior.  The increased temperature leads to y$^{+}$ = 0.0802 (at $\chi^2$ = 0.05), but this is only 0.3\% high compared with an uncertainty of nearly 5\%.  This uncertainty (on y$^{+}$) itself benefits from the prior, especially on the lower side, as would be expected due to $T_{OIII} > T_{HeII}$.  In effect, the inclusion of the [O~III] temperature prior, via our conservative implementation, produces a negligible bias, reforms parameter behavior, and strengthens the determination of the solution.

\begin{table}
\centering
\vskip .1in
\begin{tabular}{lcccc}
\hline\hline
				& Input	 & MCMC					& MCMC w/ prior \\
\hline
T(O~III)			& 19,000 &					&				\\
He$^+$/H$^+$			& 0.08 	 & 0.07999 $^{+0.00433}_{-0.00285}$ 	& 0.08021 $^{+0.00362}_{-0.00236}$ \\
n$_e$				& 100.0  & 100.2 $^{+   209.2}_{-   100.2}$	& 89.5 $^{+   126.1}_{-    89.5}$ \\
ABS(He~I)			& 1.0	 & 1.00 $^{+ 0.09}_{- 0.05}$		& 1.00 $^{+ 0.08}_{- 0.05}$ \\
$\tau$				& 2.0    & 2.00 $^{+ 0.57}_{- 0.49}$		& 1.96 $^{+ 0.54}_{- 0.40}$ \\
T$_e$				& 18,000 & 17994 $^{+2228}_{-2731}$ 		& 18,333 $^{+1700}_{-2053}$ \\
C(H$\beta$)			& 0.1    & 0.10 $^{+ 0.02}_{- 0.03}$		& 0.10 $^{+ 0.01}_{- 0.03}$ \\
ABS(H~I)			& 1.0    & 1.00 $^{+ 0.83}_{- 0.59}$		& 1.01 $^{+ 0.76}_{- 0.59}$ \\
$\xi$ $\times$ 10$^4$   	& 1.0    & 1.00 $^{+    61.50}_{-     1.00}$	& 0.38 $^{+    46.49}_{-     0.38}$ \\
\hline
\end{tabular}
\caption{Comparison of the effect of a T(O~III) prior with Synthetic Data}
\label{table:Syn_TOIII}
\end{table}

\section{Application of MCMC to real observations} \label{Galaxies}

To illustrate the effect of using the MCMC method, we will calculate the resulting value of the helium abundance y$^{+}$, along with the other input parameters and their associated uncertainties, for the seven galaxies from \citet{it98} analyzed in ref.~\cite{os04} and AOS.  Additionally, NGC~346 \citep{ppr00} and I~Zw~18SE \citep{izo99} are also included, as in AOS.  The values of T(O~III) used are taken from ref.~\cite{os04}.

MCMC, with the T(O~III) prior, was used to determine the likelihood function for each of the nine objects considered in AOS.  The best fit value of the input parameters and their uncertainties are found from the likelihood function as described in section \S \ref{Synthetic}.  Mrk~193, SBS~1420+544, and SBS~0335-052E each exhibit a pronounced double minimum.  As expected from the synthetic testing in \S \ref{Tau}, these three objects have the largest optical depths.  

\begin{figure}
\centering  
\resizebox{\textwidth}{!}{\includegraphics{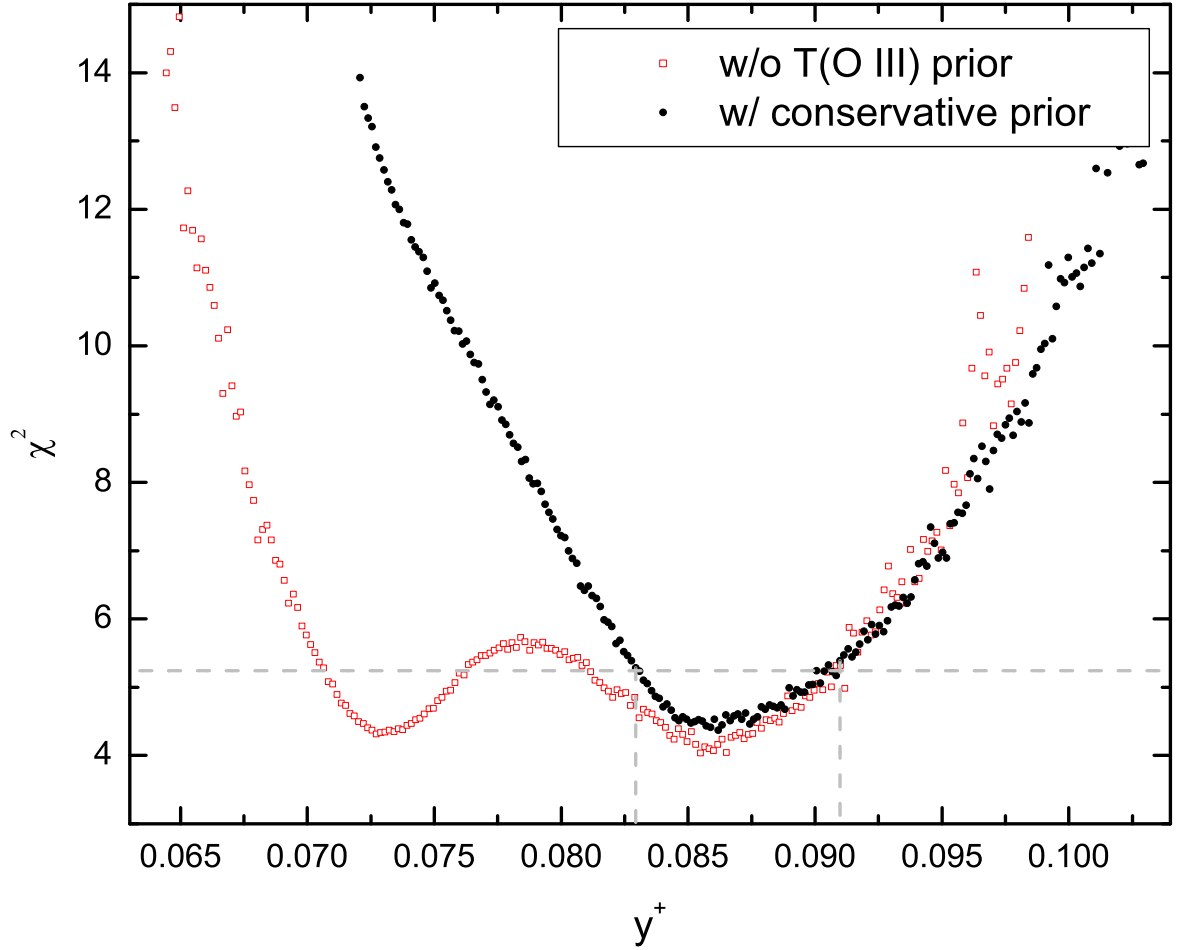}}
\caption{
$\chi^{2}$ versus abundance for Mrk~193.  The darker, solid circles are analyzed with the T(O~III) prior, while the lighter, open squares are without.   The 68\% confidence level is marked.  The prominent double minima, though similar in $\chi^2$, correspond to very different temperatures and, therefore, are readily resolved by including the T(O~III) prior.  The shift in the minimum is negligible in comparison to its uncertainty.
}
\label{193-y}
\end{figure}

As an example of the degeneracies for the large optical depth regime, we examine Mrk~193.  Shown in figure \ref{193-y}, the T(O~III) prior clearly distinguishes the physical from the unphysical minimum.  Before including the O~III measurement, the two minima are troublingly similar in their $\chi^2$ values, yet dramatically different in their physical environment.  The slightly higher minimum with $\chi^2=4.3$, has a lower abundance, lower temperature, and higher density than the one with $\chi^2=4.0$.  The parameter values of the higher minimum are y$^{+}$ = 0.07271, n$_{e}$ = 5675 $cm^{-3}$, and T = 6,655 K.  These values of density and temperature are physically unreasonable.  The lower minimum has values, y$^{+}$ = 0.08572, n$_{e}$ = 2 $cm^{-3}$, and T = 14,017 K, which are certainly quite reasonable.  Although each minimum defines a 68\% CL by itself, neither does so at  the 95\% CL.  Therefore, even though the lower minimum is the physically relevant one, their near equivalence in $\chi^2$ and poorly constrained uncertainty values would undermine the reliability of this object.  By including the [O~III] temperature ($T_{OIII}=15,561~K$), the unphysical minimum is completely removed, and the galaxy produces a well-defined solution: y$^{+}$ = 0.08619, n$_{e}$ = 1 $cm^{-3}$, and T = 15,226 K.  This new minimum is shifted 0.5\% higher, but this variance is insignificant in comparison with the overall uncertainty of 5\%.  Unlike in the case of synthetic testing, the difference in the solutions with and without the T(O~III) prior is not a measure of the bias itself, but of an upper bound on the bias.  Here, the solution excluding the [O~III] measurement is not necessarily more accurate and, in fact, can be seen as less accurate due to its neglect of a relevant observation and larger uncertainty.

\begin{figure}
\centering  
\resizebox{\textwidth}{!}{\includegraphics{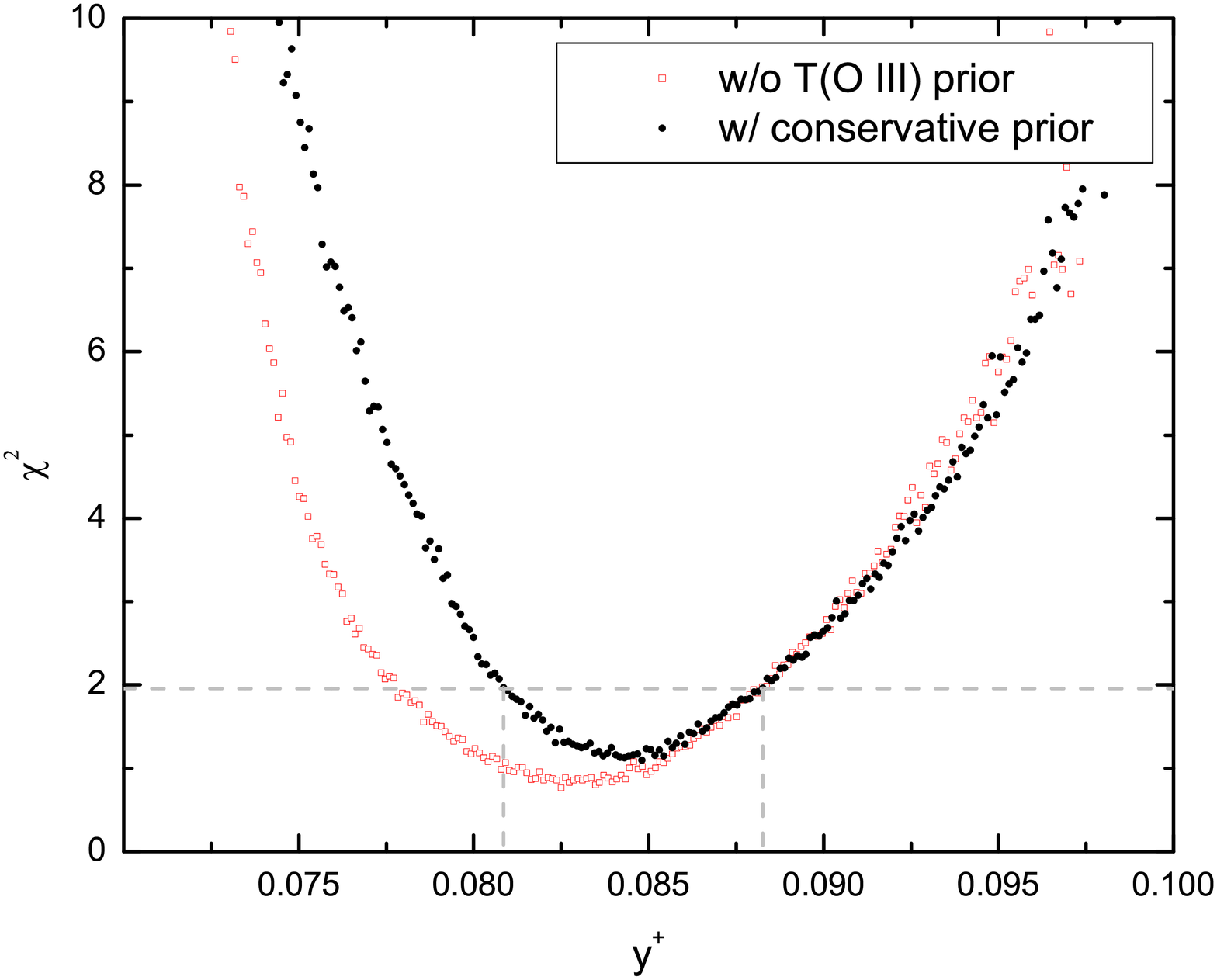}}
\caption{
$\chi^{2}$ versus abundance for SBS~1159+545.  The darker, solid circles are analyzed with the T(O~III) prior, while the lighter, open squares are without.  The 68\% confidence level marked.  Very similar to synthetic data (figure \ref{Syn-y}), this object is characteristic of well-behaved parameter determinations.  The notable shift in the minimum is due to a large difference in T(O~III) and T(He~II), but is still much less than the uncertainty.
}
\label{1159-y}
\end{figure}

The remaining six objects are much better behaved.  Figure \ref{1159-y} exhibits the solution for one of these, SBS~1159+545.  For these objects the shapes of the likelihood plots are similar to those found in the synthetic results of figures \ref{Syn-y}, \ref{Syn-He_4panel}, and \ref{Syn-H_3panel}, prior to the inclusion of the [O~III] prior. For SBS~1159+545, the T(O~III) prior noticeably increases the helium abundance found in the solution.  This occurs primarily due to the higher temperatures found when the prior is included 
(using the T$_{OIII}$ value of 18,600 K for SBS 1159+545 raises the solution temperature, T, from 15,900 K to 17,100 K).  Reassuringly, this large temperature difference only raises the abundance by 1\% (0.08326 to 0.08416), compared with an uncertainty of 5\%.  As was previously mentioned in discussing Mrk~193, the solution with T(O~III) is preferred because the added information strengthens the physical determination and reliability, outweighing the small unwanted bias.

\begin{figure}
\centering  
\resizebox{\textwidth}{!}{\includegraphics{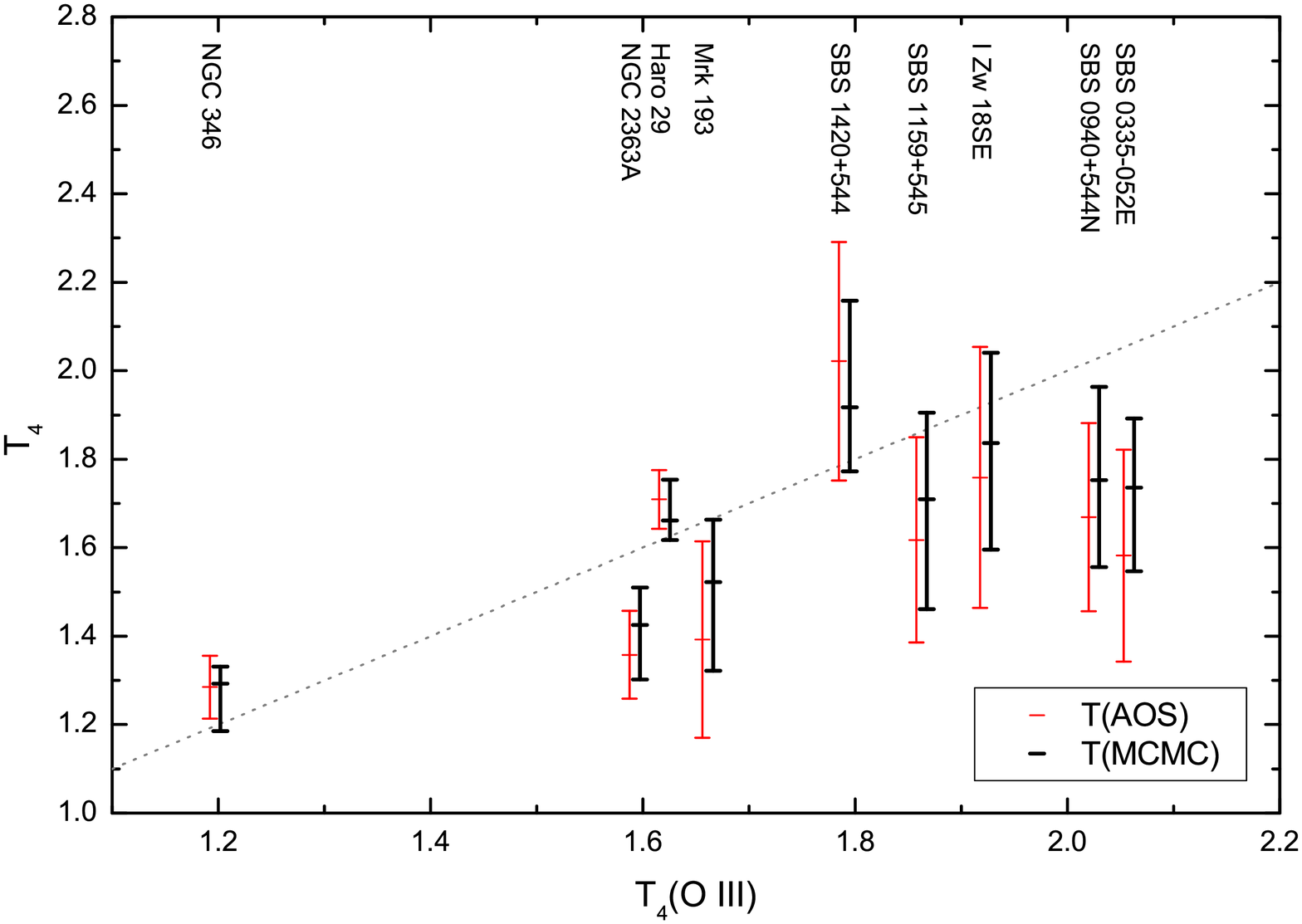}}
\caption{
Plot of the solved temperatures versus T(O~III).  The lighter, thinner lines show the results given in AOS; while our MCMC (utilizing the T(O~III) prior) temperatures are given in darker, thicker bars.  The straight, dotted line of slope one denotes equal solved and [O~III] temperatures.  Note the deviations from this line exceeding the uncertainty, including a cluster at higher temperature.
}
\label{TvTOIII}
\end{figure}

Figure \ref{TvTOIII} is included to illustrate the importance of using the T(O~III) prior conservatively.  Though figure \ref{193-y} clearly shows the dramatic benefit of incorporating the [O~III] measurement, we chose a very conservative implementation.  Had we used a more strict T(O~III) prior, any possible deviation between T(He~II) and T(O~III) would have been washed out.  Interestingly, three objects show He temperatures {\em greater} than the T(O~III) temperature, though they are all within 1$\sigma$ of being equal. The remainder of the objects do show lower He temperatures, as might be expected, and it is significantly lower for several galaxies.  Therefore, our weak prior still allows the solved temperature to reflect the physical environment defined by the helium emission.

The results, best-fit solution and uncertainties, for the entire sample are presented in 
table \ref{table:GTO}.  These are compared with the results from AOS and are found to be 
similar (differences of less than 1 $\sigma$) for the all of the objects.  
The most significant difference in the helium abundance is for SBS~0335-052E, 
where the abundance likelihood has a very broad, shallow bottom and a weak second minimum.  
As a result, the previous solution for SBS~0335-052E was determined by an average of 
abundances that spanned this large range in y$^{+}$, including the weak secondary 
minimum with a low abundance and temperature.  Therefore, the new solution, consistent 
with the T(O~III) prior, is much higher. In the three cases with the largest shifts in
helium abundance, SBS~0335-052E, SBS~0940+544N, and NGC~2363A, the helium abundance
increased due to an increase in the temperature, which results from the use of the T(O~III) prior.

\begin{landscape}
\begin{table}[b!]\footnotesize
\centering
\vskip .1in
\begin{tabular}{lccccccccc}
\hline
\hline
Object 			& He$^+$/H$^+$     		   &  n$_e$     		   & ABS(He~I)     		& $\tau$     				& T$_e$     		    & T(O~III) & C(H$\beta$)     	  & ABS(H~I)     	       & $\xi$ $\times$ 10$^4$ \\
\hline
\hline
&&&& AOS \\
\hline
I~Zw~18SE & 0.08008 $\pm$ 0.00469 &     77 $\pm$    277 &  0.42 $\pm$  0.36 &  0.42 $\pm$  0.56 & 17,590 $\pm$ 2950 & - &  0.001 $\pm$  0.005 &  4.76 $\pm$  0.82 &      8 $\pm$     70  \\
SBS~0335-052E & 0.08536 $\pm$ 0.00644 &    121 $\pm$    254 &  0.39 $\pm$  0.31 &  5.32 $\pm$  0.87 & 15,820 $\pm$ 2400 & - &  0.052 $\pm$  0.045 &  1.82 $\pm$  1.04 &    117 $\pm$    262  \\
SBS~0940+544N & 0.08506 $\pm$ 0.00484 &    149 $\pm$    277 &  0.09 $\pm$  0.20 &  0.61 $\pm$  0.56 & 16,690 $\pm$ 2120 & - &  0.000 $\pm$  0.000 &  0.03 $\pm$  0.17 &     26 $\pm$     33  \\
SBS~1159+545 & 0.08332 $\pm$ 0.00390 &    325 $\pm$    330 &  0.03 $\pm$  0.06 &  0.39 $\pm$  0.43 & 16,180 $\pm$ 2320 & - &  0.021 $\pm$  0.023 &  0.59 $\pm$  0.63 &     68 $\pm$    152  \\
SBS~1420+544 & 0.08939 $\pm$ 0.00393 &     83 $\pm$    234 &  0.18 $\pm$  0.16 &  2.89 $\pm$  0.62 & 20,210 $\pm$ 2700 & - &  0.156 $\pm$  0.028 &  0.02 $\pm$  0.14 &     30 $\pm$    200  \\
Haro~29 & 0.08576 $\pm$ 0.00174 &     37 $\pm$     48 &  0.52 $\pm$  0.15 &  0.13 $\pm$  0.16 & 17,100 $\pm$  670 & - &  0.000 $\pm$  0.000 &  3.38 $\pm$  0.24 &      0.2 $\pm$      0.4  \\
Mrk~193 & 0.08660 $\pm$ 0.00435 &    117 $\pm$    320 &  0.45 $\pm$  0.19 &  2.63 $\pm$  0.78 & 13,920 $\pm$ 2220 & - &  0.213 $\pm$  0.043 &  0.64 $\pm$  0.74 &   1003 $\pm$   8466  \\
NGC~2363A & 0.08328 $\pm$ 0.00244 &    194 $\pm$    203 &  0.46 $\pm$  0.18 &  1.55 $\pm$  0.34 & 13,580 $\pm$  1000 & - &  0.117 $\pm$  0.010 &  0.29 $\pm$  0.44 &      5 $\pm$     49  \\
NGC~346 & 0.08777 $\pm$ 0.00168 &     30 $\pm$     70 &  0.37 $\pm$  0.15 &  0.03 $\pm$  0.09 & 12,850 $\pm$  710 & - &  0.132 $\pm$  0.013 &  0.10 $\pm$  0.28 &    359 $\pm$    758  \\
\hline
&&&& MCMC \\
&&&& Re-Analysis \\
\hline
I~Zw~18SE         	& 0.08102 $^{+0.00323}_{-0.00540}$ &      1 $^{+   179}_{-     1}$ &  0.35 $^{+ 0.21}_{- 0.24}$ &  0.38 $^{+ 0.62}_{- 0.38}$ & 18,360 $^{+2050}_{-2410}$ & 19,180 &  0.00 $^{+ 0.02}_{- 0.00}$ &  5.01 $^{+ 0.62}_{- 0.89}$ &      0 $^{+    14}_{-     0}$ \\
SBS~0335-052E        	& 0.08991 $^{+0.00715}_{-0.00843}$ &      1 $^{+   130}_{-     1}$ &  0.40 $^{+ 0.22}_{- 0.23}$ &  5.24 $^{+ 0.64}_{- 0.69}$ & 17,360 $^{+1560}_{-1900}$ & 20,530 &  0.06 $^{+ 0.06}_{- 0.05}$ &  1.92 $^{+ 1.13}_{- 1.22}$ &     46 $^{+   123}_{-    46}$ \\
SBS~0940+544N        	& 0.08750 $^{+0.00488}_{-0.00498}$ &     70 $^{+   230}_{-    70}$ &  0.00 $^{+ 0.19}_{- 0.00}$ &  0.66 $^{+ 0.57}_{- 0.66}$ & 17,520 $^{+2110}_{-1960}$ & 20,200 &  0.00 $^{+ 0.01}_{- 0.00}$ &  0.00 $^{+ 0.33}_{- 0.00}$ &     30 $^{+    45}_{-    23}$ \\
SBS~1159+545       	& 0.08416 $^{+0.00410}_{-0.00331}$ &    213 $^{+   252}_{-   138}$ &  0.00 $^{+ 0.08}_{- 0.00}$ &  0.18 $^{+ 0.58}_{- 0.18}$ & 17,100 $^{+1960}_{-2490}$ & 18,580 &  0.04 $^{+ 0.03}_{- 0.04}$ &  0.40 $^{+ 0.95}_{- 0.40}$ &     22 $^{+   133}_{-    22}$ \\
SBS~1420+544       	& 0.09058 $^{+0.00435}_{-0.00498}$ &     52 $^{+   138}_{-    52}$ &  0.15 $^{+ 0.12}_{- 0.15}$ &  3.13 $^{+ 0.42}_{- 0.66}$ & 19,180 $^{+2400}_{-1450}$ & 17,850 &  0.15 $^{+ 0.03}_{- 0.02}$ &  0.00 $^{+ 0.26}_{- 0.00}$ &     13 $^{+    27}_{-    13}$ \\
Haro~29         	& 0.08472 $^{+0.00225}_{-0.00159}$ &     56 $^{+    61}_{-    56}$ &  0.45 $^{+ 0.11}_{- 0.11}$ &  0.14 $^{+ 0.25}_{- 0.14}$ & 16,610 $^{+ 930}_{- 440}$ & 16,150 &  0.00 $^{+ 0.00}_{- 0.00}$ &  3.22 $^{+ 0.20}_{- 0.24}$ &      0 $^{+     1}_{-     0}$ \\
Mrk~193        		& 0.08619 $^{+0.00480}_{-0.00326}$ &      1 $^{+   137}_{-     1}$ &  0.38 $^{+ 0.12}_{- 0.11}$ &  2.40 $^{+ 0.70}_{- 0.44}$ & 15,230 $^{+1410}_{-2010}$ & 16,560 &  0.25 $^{+ 0.02}_{- 0.05}$ &  0.26 $^{+ 0.78}_{- 0.26}$ &     22 $^{+   357}_{-    22}$ \\
NGC~2363A       	& 0.08469 $^{+0.00197}_{-0.00300}$ &     73 $^{+   204}_{-    73}$ &  0.50 $^{+ 0.14}_{- 0.15}$ &  1.71 $^{+ 0.20}_{- 0.29}$ & 14,260 $^{+ 850}_{-1230}$ & 15,870 &  0.12 $^{+ 0.01}_{- 0.01}$ &  0.11 $^{+ 0.30}_{- 0.11}$ &      0 $^{+    25}_{-     0}$ \\
NGC~346        		& 0.08803 $^{+0.00125}_{-0.00179}$ &     15 $^{+   104}_{-    15}$ &  0.27 $^{+ 0.08}_{- 0.08}$ &  0.00 $^{+ 0.14}_{- 0.00}$ & 12,930 $^{+ 380}_{-1080}$ & 11,190 &  0.14 $^{+ 0.01}_{- 0.01}$ &  0.00 $^{+ 0.17}_{- 0.00}$ &    516 $^{+  1124}_{-   204}$ \\
\hline
\end{tabular}
\caption{Comparison of Physical Conditions and He$^+$/H$^+$ Abundance Solutions}
\label{table:GTO}
\end{table}
\end{landscape}

\section{Results of the MCMC analysis} \label{Results}

\begin{figure}
\centering  
\resizebox{\textwidth}{!}{\includegraphics{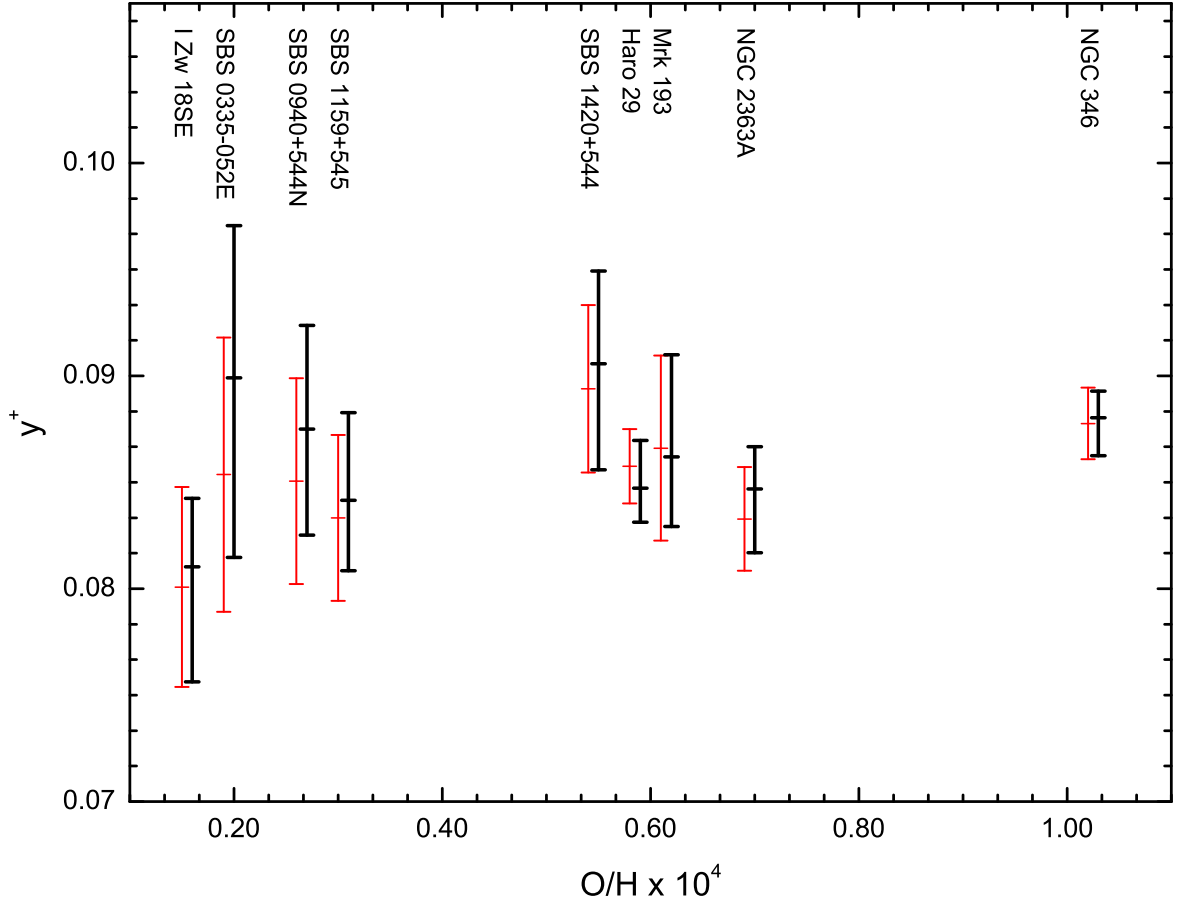}}
\caption{
Abundance comparison for the target objects as analyzed in AOS, using Gaussian distributed fluxes, and MCMC.  The lighter, thinner lines are for Gaussian fluxes, with MCMC given in thicker, darker bars.  The quoted primordial helium abundance, Y$_{p}$, is based on a regression of the seven objects of AOS (i.e., I~Zw~18SE and NGC~346 have not been used in the regression).  This sample is also used to produce the mean helium abundance, $<Y>$.
}
\label{y_PIII_MCMC}
\end{figure}

A comparison of the derived values of y$^{+}$ presented in table \ref{table:GTO} is plotted 
in figure \ref{y_PIII_MCMC}.  An important result of this comparison, between the previous 
analysis of AOS and the MCMC method, is that the uncertainties are very similar.  Indeed, the 
maximum uncertainty (the larger of the two asymmetric uncertainties) is larger but, in all 
cases, is much less than 1 $\sigma$ larger.  On average, the maximum uncertainty increased by 17\%.  

That the uncertainties increased at all might be unexpected.  Naively, one would
expect that the use of the conservative prior, which eliminates an unphysical 
minimum, would result in a decrease in the uncertainty.  However, the comparison between 
the analysis in AOS and the MCMC method with prior reveals that the previous method inherently
underestimated the errors.  For example, for Mrk~193 (figure \ref{193-y}), the available
parameter space for the Monte Carlo analysis was much broader for the AOS analysis, but the 
estimate of the uncertainty did not capture the true uncertainty.  Primarily, this is because 
the variance was calculated symmetrically about the mean, not asymmetrically about the minimum, 
and because the method was not a rigorous approach to exploring the parameter variation.
The MCMC method yields an improved statistical result and more accurately reflects the 
uncertainty of the helium abundance determination.

The development of the new MCMC analysis technique is the primary goal of this work.  To aid in demonstrating the effect, we also calculate the primordial helium abundance (mass fraction), Y$_{p}$.  A regression of Y, the helium mass fraction, versus O/H, the oxygen to hydrogen mass fraction, is used to extrapolate to the primordial value.\footnote{This work takes $Z=20(O/H)$ such that $Y=\frac{4y(1-20(O/H))}{1+4y}$}  For direct comparison, the O/H values are taken from AOS \citep[which took the values from][for the same reason]{os04}.  The relevant values are listed in table \ref{table:PH}.  

NGC~346 is more chemically evolved, and I~Zw~18SE has a low equivalent width of H$\beta$, making it more sensitive to underlying absorption, and is a candidate for Galactic Na~I absorption due to its radial velocity.  As a result, neither object was included in the primary regression of AOS and will similarly be excluded here.  This regression yields $Y_p = 0.2609 \pm 0.0117$,
with slope -67 $\pm$ 214 and $\chi^{2}$ = 1.7.  Note that $dY/dZ$ is very poorly determined and 
if we restrict the analysis to theoretically meaningful positive slopes, 
we find 
\beq
Y_p = 0.2573^{+0.0033}_{-0.0088} ,
\label{eq:Yp}
\eeq
with a range in slopes from 0 to 149 and a $\chi^2$ of 1.8.  Equation \ref{eq:Yp} agrees with the WMAP value of $Y_p = 0.2487 \pm 0.0002$ within 1$\sigma$.  We do note, however, that the two most recent measurements of the neutron mean life \citep{ser,pich} are significantly lower than the accepted world average \citep{rpp}. A drop in the mean life of about 6 s would result in a lower BBN abundance by about 0.001.  While, conservatively, it is premature to claim a discrepancy in helium (because the systematic uncertainties are so large), if real, it may hint towards new physics \citep{ham,kra,naka} or require an early astrophysical source for helium \citep{vsof}. AOS determined $Y_p = 0.2561 \pm 0.0108$ ($0.2561^{+0.0032}_{-0.0108}$ when the slope is restricted to be positive); that these results agree so closely is not surprising given that the dataset and physical model of the works are identical.  

As the O/H domain is very limited, a mean evaluation is justified and gives,
\beq
Y_p = 0.2573 \pm 0.0033.
\eeq
Inclusion of I~Zw~18SE and NGC~346, the lowest and highest metallicity objects, reduces the intercept and error to $Y_p = 0.2549 \pm 0.0066$ with a slope of 45 $\pm$ 86.  Much of the improvement stems from the longer metallicity baseline but, therefore, also more assumptions.  It has always been our view that one well measured object at high metallicity, which would fix $dY/dZ$, should not affect the uncertainties of  measurements at low metallicities if their uncertainties are large.  The size of our overall uncertainty also suffers from a low number of ``high quality'' measurements at low metallicity which should be close to primordial.  Nonetheless, this is the state of current data. Including more ``high quality'' objects would clearly benefit the determination.  This work has focused on introducing a new statistical method, saving a revised sample for later work.  Indeed, we hope that this method will allow the unambiguous use of other low metallicity data which are found in the literature.

\begin{table}[ht!]
\centering
\vskip .1in
\begin{tabular}{lcccc}
\hline\hline
Object & 	He$^+$/H$^+$ 	      & He$^{++}$/H$^+$     & Y 		  & O/H $\times$ 10$^4$ \\
\hline
I~Zw~18 & 	0.08102	$\pm$ 0.00540 & 0.0024 $\pm$ 0.0024 & 0.2501 $\pm$ 0.0133 & 0.15 $\pm$ 0.01 \\
SBS~0335-052 & 	0.08991	$\pm$ 0.00843 & 0.0023 $\pm$ 0.0023 & 0.2694 $\pm$ 0.0187 & 0.19 $\pm$ 0.01 \\ 
SBS~0940+544N & 0.08750	$\pm$ 0.00498 & 0.0000 		    & 0.2591 $\pm$ 0.0109 & 0.26 $\pm$ 0.01 \\ 
SBS~1159+545 & 	0.08416	$\pm$ 0.00410 & 0.0010 $\pm$ 0.0010 & 0.2539 $\pm$ 0.0094 & 0.30 $\pm$ 0.01 \\ 
SBS~1420+544 & 	0.09058	$\pm$ 0.00498 & 0.0011 $\pm$ 0.0011 & 0.2680 $\pm$ 0.0109 & 0.54 $\pm$ 0.01 \\ 
Haro~29 & 	0.08472	$\pm$ 0.00225 & 0.0011 $\pm$ 0.0011 & 0.2552 $\pm$ 0.0055 & 0.58 $\pm$ 0.01 \\ 
Mrk~193 &       0.08619	$\pm$ 0.00480 & 0.0011 $\pm$ 0.0011 & 0.2584 $\pm$ 0.0108 & 0.61 $\pm$ 0.02 \\
NGC~2363A & 	0.08469	$\pm$ 0.00300 & 0.0012 $\pm$ 0.0012 & 0.2554 $\pm$ 0.0072 & 0.69 $\pm$ 0.01 \\
NGC~346 & 	0.08803	$\pm$ 0.00179 & 0.0000 		    & 0.2599 $\pm$ 0.0039 & 1.02 $\pm$ 0.02 \\
\hline
\end{tabular}
\caption{Primordial Helium Regression Values}
\label{table:PH}
\end{table}

\section{Summary} \label{Conclusion}

The primary result of this work is the demonstration of a statistically rigorous method for determining the uncertainty of the abundance and model parameters.  The use of MCMC allows one to efficiently sample the parameter space so that the uncertainties can be calculated directly from the change in $\chi^{2}$ as the parameters are varied from the best-fit solution.  Computationally, MCMC is efficient and straightforward.  Beyond the improvement in the approach itself, the constructed likelihood distributions for the parameters are instructive in evaluating the quality of the object and illuminating the sources of differences with previous analyses.

Particularly illustrative of the benefits of the likelihood approach was the discovery of the increased degeneracy at large optical depth.  With synthetic data, a second minimum emerges as the optical depth increases.  As this false minimum becomes more significant, the reliability and quality of the object is undermined.  In concordance with this predicted effect, several galaxies, each with large optical depth, exhibit prominent second minimums.  However, a second benefit of the approach is the straightforward incorporation and interpretation of priors.  This allows the electron temperature defined by the [O~III] emission lines to be utilized to eliminate low temperature, unphysical secondary minimums.  Therefore, after taking the [O~III] measurement into account, large optical depth objects are well behaved.  It also worthy of note that the prior is used very conservatively.  This ensures that the solved value of the temperature primarily reflects the helium defined temperature, thus protecting the abundance from any significant bias.  

Thus, the new MCMC method is a distinct improvement, resulting in a statistically more accurate determination of the helium abundance, the physical parameters associated with the HII region, and their uncertainties. Nevertheless, we found relatively large uncertainties  in the helium abundance determinations of individual low metallicity HII regions.  This, however, is an indication of the true uncertainty in the measurement and the challenge posed.   Future work will investigate the possibilities for improving the result through the use of a revised and expanded set of objects.

\acknowledgments

The work of EA and KAO is supported in part by DOE grant DE-FG02-94ER-40823. EDS is grateful for partial support from the University of Minnesota.

\end{document}